# OLTP on Hardware Islands

Danica Porobic    Ippokratis Pandis[†]    Miguel Branco    Pınar Tözün    Anastasia Ailamaki

École Polytechnique Fédérale de Lausanne
Lausanne, VD, Switzerland
{danica.porobic, miguel.branco, pinar.tozun, anastasia.ailamaki}@epfl.ch

[†]IBM Almaden Research Center
San Jose, CA, USA
ipandis@us.ibm.com

## ABSTRACT

Modern hardware is abundantly parallel and increasingly heterogeneous. The numerous processing cores have non-uniform access latencies to the main memory and to the processor caches, which causes variability in the communication costs. Unfortunately, database systems mostly assume that all processing cores are the same and that microarchitecture differences are not significant enough to appear in critical database execution paths. As we demonstrate in this paper, however, hardware heterogeneity does appear in the critical path and conventional database architectures achieve suboptimal and even worse, unpredictable performance.

We perform a detailed performance analysis of OLTP deployments in servers with multiple cores per CPU (*multicore*) and multiple CPUs per server (*multisocket*). We compare different database deployment strategies where we vary the number and size of independent database instances running on a single server, from a single shared-everything instance to fine-grained shared-nothing configurations. We quantify the impact of non-uniform hardware on various deployments by (a) examining how efficiently each deployment uses the available hardware resources and (b) measuring the impact of distributed transactions and skewed requests on different workloads. Finally, we argue in favor of shared-nothing deployments that are topology- and workload-aware and take advantage of fast on-chip communication between *islands* of cores on the same socket.

## 1. INTRODUCTION

On-Line Transaction Processing (OLTP) is a multi-billion dollar industry[1] and one of the most important and demanding database applications. Innovations in OLTP continue to deserve significant attention, advocated by the recent emergence of appliances[2], startups[3], and research projects (e.g. [31, 18, 25, 21, 24]). OLTP applications are mission-critical for many enterprises with little margin for compromising either performance or scalability. Thus, it is not surprising that all major OLTP vendors spend significant effort in developing highly-optimized software releases, often with platform-specific optimizations.

Over the past decades, OLTP systems benefited greatly from improvements in the underlying hardware. Innovations in their software architecture have been plentiful but there is a clear benefit from processor evolution. Uni-processors grew predictably faster with time, leading to better OLTP performance. Around 2005, when processor vendors hit the frequency-scaling wall, they started obtaining performance improvements by adding multiple processing cores to the same CPU chip, forming chip multiprocessors (multicore or CMP); and building servers with multiple CPU sockets of multicore processors (SMP of CMP).

Multisockets of multicores are highly parallel and characterized by heterogeneity in the communication costs: sets, or *islands*, of processing cores communicate with each other very efficiently through common on-chip caches, and communicate less efficiently with others through bandwidth-limited and higher-latency links. Even though multisocket multicore machines dominate in modern data-centers, it is unclear how well software systems and in particular OLTP systems exploit hardware capabilities.

This paper characterizes the impact of hardware topology on the behavior of OLTP systems running on modern multisocket multicore servers. As recent studies argue and this paper corroborates, traditional shared-everything OLTP systems underperform on modern hardware because of (a) excessive communication between the various threads [5, 14] and (b) contention among threads [26, 31]. Practitioners report that even commercial shared-everything systems with support for non-uniform memory architectures (NUMA) underperform [11, 36]. On the other hand, shared-nothing deployments [30] face the challenges of (a) higher execution costs when distributed transactions are required [16, 9, 12, 27], even within a single node, particularly if the communication occurs between slower links (e.g. across CPU sockets); and (b) load imbalances due to skew [33].

Many real-life workloads cannot be easily partitioned across instances or can have significant data and request skews, which may also change over time. In this paper, we examine the impact of perfectly partitionable and non-partitionable workloads, with and without data skew, on shared-nothing deployments of varying sizes as well as shared-everything deployments. Our experiments show that per-fectly partitionable workloads perform significantly better on fine-grained shared-nothing configurations but non-partitionable workloads favor coarse-grained configurations, due to the overhead of distributed transactions. We identify the overheads as messaging, additional logging, and increased contention, all of which depend on workload characteristics such as the percentage of multisite transactions, the number of sites touched by each transaction, and the amount of work done within each transaction. Additionally, we find that skewed accesses cause performance to drop significantly when using fine-grained shared-nothing configurations; this effect is less evident on coarser configurations and when using shared-everything deployments.

---

[1]E.g. http://www.gartner.com/DisplayDocument?id=1044912
[2]Such as Oracle's Exadata database machine.
[3]Such as VoltDB, MongoDB, NuoDB, and others.

To our knowledge, this is the first study that systematically analyzes the performance of shared-everything and shared-nothing OLTP configurations of varying size on modern multisocket multicore machines. The contributions are as follows:

- We provide experimental evidence of the impact of non-uniform hardware on the performance of transaction processing systems and conclude that high performance software has to minimize contention among cores and avoid frequent communication between distant cores.

- Our experiments show that fine-grained shared-nothing deployments can achieve more than four times as high throughput as a shared-everything system when the workload is perfectly partitionable. By contrast, when the workload is not partitionable and/or exhibits skew, shared-everything achieves twice as high a throughput as shared-nothing. Therefore, there is no unique optimal deployment strategy that is independent of the workload.

- We demonstrate that a careful assignment of threads to islands of cores can combine the best features of a broad range of system configurations, thereby achieving flexibility in the deployment as well as more predictable and robust performance. In particular, islands-aware thread assignment can improve the worst-case scenario by a factor of 2 without hurting the best-case performance much.

The rest of the document is structured as follows. Section 2 presents the background and related work, describing the two main database deployment approaches. Section 3 identifies recent trends on modern hardware and their implications on software design. Section 4 discusses the dependence of database systems performance on hardware topology and workload characteristics such as percentage of distributed transactions. Section 5 presents experimental methodology. Section 6 describes cases favoring fine-grained shared-nothing configurations, and Section 7 analyzes possible overheads when deploying shared-nothing configurations. Finally, Section 8 summarizes the findings and discusses future work.

## 2. BACKGROUND AND RELATED WORK

Shared-everything and shared-nothing database designs, described in the next two sections, are the most widely used approaches for OLTP deployments. Legacy multisocket machines, which gained popularity in the 1990s as symmetric multiprocessing servers, had non-uniform memory access (NUMA) latencies. This required changes to the database and operating systems to diminish the impact of NUMA, as discussed in Section 2.3.

### 2.1 Shared-everything Database Deployments

Within a database node, *shared-everything* is any deployment where a single database instance manages all the available resources. As database servers have long been designed to operate on machines with multiple processors, shared-everything deployments assume equally fast communication channels between all processors, since each thread needs to exchange data with its peers. Until recently, shared-everything was the most popular deployment strategy on a single node. All major commercial database systems adopt it.

OLTP has been studied extensively on shared-everything databases. For instance, transactions suffer significant stalls during execution [3, 2, 14]; a result we corroborate in Section 6.2. It has also been shown that shared-everything systems have frequent shared read-write accesses [5, 14], which are difficult to predict [29]. Modern systems enter numerous contentious critical sections even when executing simple transactions, affecting single-thread performance, requiring frequent inter-core communication, and causing contention among threads [26, 25, 18]. These characteristics make distributed memories (as those of multisockets), distributed caches (as those of multicores), and prefetchers ineffective. Recent work suggests a departure from the traditional transaction-oriented execution model, to adopt a data-oriented execution model, circumventing the aforementioned properties - and flaws - of traditional shared-everything OLTP [25, 26].

### 2.2 Shared-nothing Database Deployments

*Shared-nothing* deployments [30], based on fully independent (physically partitioned) database instances that collectively process the workload, are an increasingly appealing design even within single node [31, 21, 28]. This is due to the scalability limitations of shared-everything systems, which suffer from contention when concurrent threads attempt to access shared resources [18, 25, 26].

The main advantage of shared-nothing deployments is the explicit control over the contention within each physical database instance. As a result, shared-nothing systems exhibit high single-thread performance and low contention. In addition, shared-nothing databases typically make better use of the available hardware resources whenever the workload executes transactions touching data on a single database instance. Systems such as H-Store [31] and HyPer [21] apply the shared-nothing design to the extreme, deploying one single-threaded database instances per CPU core. This enables simplifications or removal of expensive database components such as logging and locking.

Shared-nothing systems appear ideal from the hardware utilization perspective, but they are sensitive to the ability to partition the workload. Unfortunately, many workloads are not perfectly partitionable, i.e. it is hardly possible to allocate data such that every transaction touches a single instance. Whenever multiple instances must collectively process a request, shared-nothing databases require expensive distributed consensus protocols, such as two-phase commit, which many argue are inherently non-scalable [16, 9]. Similarly, handling data and access skew is problematic [33].

The overhead of distributed transactions urged system designers to explore partitioning techniques that reduce the frequency of distributed transactions [12, 27], and to explore alternative concurrency control mechanisms, such as speculative locking [20], multiversioning [6] and optimistic

1448

concurrency control [22, 24], to reduce the overheads when distributed transactions cannot be avoided. Designers of large-scale systems have circumvented problems with distributed transactions by using relaxed consistency models such as eventual consistency [35]. Eventual consistency eliminates the need for synchronous distributed transactions, but it makes programming transactional applications harder, with consistency checks left to the application layer.

The emergence of multisocket multicore hardware adds further complexity to the on-going debate between shared-everything and shared-nothing OLTP designs. As Section 3 describes, multisocket multicores introduce an additional level into the memory hierarchy. Communication between processors is no longer uniform: cores that share caches communicate differently from cores in the same socket and other sockets.

### 2.3 Performance on Multisocket Multicores

Past work focused on adapting databases for legacy multi-socket systems. For instance, commercial database systems provide configuration options to enable NUMA support, but this setting is optimized for legacy hardware where each individual CPU is assumed to contain a single core. With newer multisocket servers, enabling NUMA support might lead to high CPU usage and degraded performance [11, 36].

An alternative approach is taken by the Multimed project, which views the multisocket multicore system as a cluster of machines [28]. Multimed uses replication techniques and a middleware layer to split database instances into those that process read-only requests and those that process updates. The authors report higher performance than a single shared-everything instance. However, Multimed does not explicitly address NUMA-awareness and the work is motivated by the fact that the shared-everything system being used has inherent scalability limitations. In this paper, we use a scalable open-source shared-everything OLTP system, Shore-MT [18], which scales nearly linearly with the available cores on single-socket machines; however, we still observe benefits with shared-nothing deployments based on Shore-MT.

A comparison of techniques for executing hash joins in multicore machines [8], corresponding broadly to shared-everything and shared-nothing configurations of different sizes, illustrates a case where shared-everything has appealing characteristics. The operation under study, however, hash joins, has different characteristics from OLTP.

Exploiting NUMA effects at the operating system level is an area of active research. Some operating system kernels such as the Mach [1] and exokernels [13], or, more recently, Barrelfish [4], employ the message-passing paradigm. Message-passing potentially facilitates the development of NUMA-aware systems since the communication between threads is done explicitly through messages, which the operating system can schedule in a NUMA-aware way. Other proposals include the development of schedulers that detect contention and react in a NUMA-aware manner [7, 32]. None of these proposals is specific to database systems and likely require extensive changes to the database engine.

### 3. HARDWARE HAS ISLANDS

Hardware has long departed from uniprocessors, which had predictable and uniform performance. Due to thermal and power limitations, vendors cannot improve the performance of processors by clocking them to higher frequency or by using more advanced techniques such as increased instruction-width and extended out-of-order execution. Instead, two approaches are mainly used to increase the processing capability of a machine. The first is to put together multiple processor chips that communicate through shared main memory. For several decades, such *multisocket* designs provided the only way to scale performance within a single node and the majority of OLTP systems have historically used such hardware. The second approach places multiple processing cores on a single chip, such that each core is capable of processing concurrently several independent instruction streams, or hardware contexts. The communication between cores in these *multicore* processors happens through on-chip caches. In recent years, multicore processors have become a commodity.

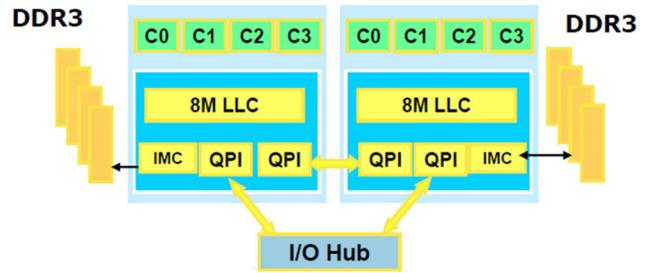

**Figure 1: Block diagram of a typical machine. Cores communicate either through a common cache, an interconnect across socket or main memory.**

Multisocket multicore systems are the predominant configuration for database servers and are expected to remain popular in the future. Figure 1 shows a simplified block diagram of a typical machine that has two sockets with quad-core CPUs.[4] Communication between the numerous cores happens through different mechanisms. For example, cores in the same socket share a common cache, while cores located in different sockets communicate via the interconnect (called *QPI* for Intel processors). Cores may also communicate through the main memory if the data is not currently cached. The result is that the inter-core communication is variable: communication in multicores is more efficient than in multisockets, which communicate over a slower, power-hungry, and often bandwidth-limited interconnect.

Hence, there are two main trends in modern hardware: the variability in communication latencies and the abundance of parallelism. In the following two subsections we discuss how each trend affects the performance of software systems.

### 3.1 Variable Communication Latencies

The impact of modern processor memory hierarchies on the application performance is significant because it causes variability in access latency and bandwidth, making the overall software performance unpredictable. Furthermore, it is difficult to implement synchronization or communication mechanisms that are globally optimal in different environments - multicores, multisockets, and multisockets of multicores.

We illustrate the problem of optimal synchronization mechanisms with a simple microbenchmark. Figure 2 plots the throughput of a program running on a machine that has 8 CPUs with 10 cores each (the "Octo-socket" machine of Table 2). There are 80 threads in the program, divided into groups of 10 threads, where each group increments a counter protected by a lock in a tight loop. There are 8 counters in total, matching the number of sockets in the machine. We vary the allocation of the worker threads and plot the total throughput (million counter increments per second). The first bar (*"Spread" threads*) spreads worker threads across all sockets. The second bar (*"Grouped" threads*) allocates all threads in the same socket as the counter. The third bar leaves the operating system to do the thread allocation. Allocating threads and memory in a topology-aware manner results in the best performance and lowest variability. Leaving the allocation to the operating system leads to non-optimal results and higher variability.[5]

---

[4]Adapted from http://software.intel.com/sites/products/collateral/hpc/vtune/performance_analysis_guide.pdf

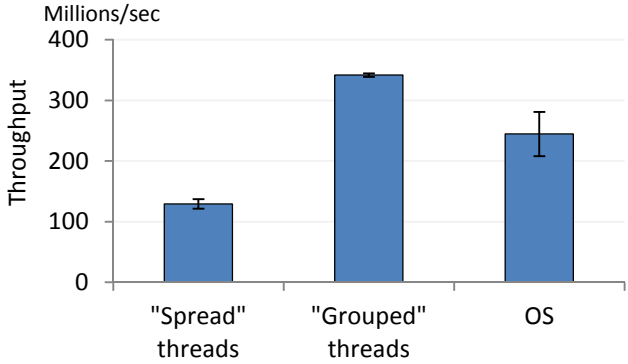

Figure 2: Allocating threads and memory in a topology-aware manner provides the best performance and lower variability.

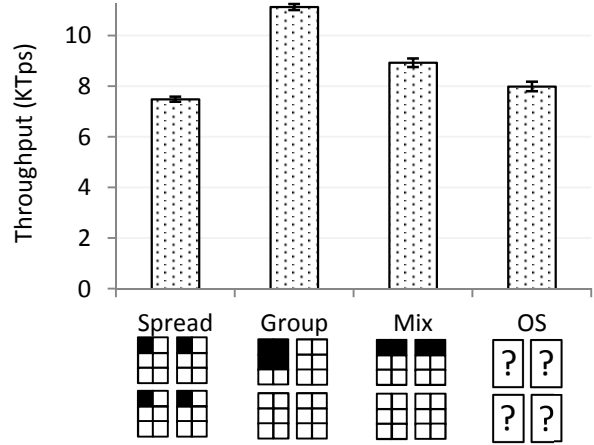

Figure 3: Running the `TPCC-Payment` workload with all cores on the same socket achieves 20-30% higher performance than other configurations.

We obtain similar results when running OLTP workloads. To demonstrate the impact of NUMA latencies on OLTP, we run `TPC-C Payment` transactions on a machine that has 4 CPUs with 6 cores each ("Quad-socket" in Table 2). Figure 3 plots the average throughput and standard deviation across multiple executions on a database with 4 worker threads. In each configuration we vary the allocation of individual worker threads to cores. The first configuration (*"Spread"*) assigns each thread to a core in a different socket. The second configuration (*"Group"*) assigns all threads to the same socket. The configuration *"Mix"* assigns two cores per socket. In the *"OS"* configuration, we let the operating system do the scheduling. This experiment corroborates the previous observations of Figure 2: the OS does not optimally allocate work to cores, and a topology-aware configuration achieves 20-30% better performance and less variation. The absolute difference in performance is much lower than in the case of counter incrementing because executing a transaction has significant start-up and finish costs, and during transaction execution a large fraction of the time is spent on operations other than accessing data. For instance, studies show that around 20% of the total instructions executed during OLTP are data loads or stores (e.g. [3, 14]).

### 3.2 Abundant Hardware Parallelism

Another major trend is the abundant hardware parallelism available in modern database servers. Higher hardware paral-lelism potentially causes additional contention in multisocket multicore systems, as a higher number of cores compete for shared data accesses. Table 1 shows the results obtained on the octo-socket machine when varying the number of worker threads accessing a set of counters, each protected by a lock. An exclusive counter per core achieves lower variability and 18x higher throughput than a counter per socket, and 517x higher throughput than a single counter for the entire machine. In both cases, this is a super-linear speedup. Shared-nothing deployments are better suited to handle contention, since they provide explicit control by physically partitioning data, leading to higher performance.

---

[5]This observation has been done also by others, e.g. [4], and is an area of active research.

Table 1: Throughput and variability when increasing counters each protected by a lock.

| Counter setup | Total counters | Throughput (M/sec) (Speedup) | Std. dev. (%) |
|---|---|---|---|
| Single | 1 | 18.4 | 9.33% |
| Per socket | 8 | 341.7 (18.5x) | 0.86% |
| Per core | 80 | 9527.8 (516.8x) | 0.03% |

In summary, modern hardware poses new challenges to software systems. Contention and topology have a significant impact on performance and predictability of the software. Predictably fast transaction processing systems have to take advantage of the *hardware islands* in the system. They need to (a) avoid frequent communication between "distant" cores in the processor topology and (b) keep the contention among cores low. The next section argues in favor of topology-aware OLTP deployments that adapt to those hardware islands.

## 4. ISLANDS: HARDWARE TOPOLOGY- AND WORKLOAD-AWARE SHARED-NOTHING OLTP

Traditionally, database systems fall into one of two main categories: shared-everything or shared-nothing. The distinction into two strict categories, however, does not capture the fact that there are many alternative shared-nothing configurations of different sizes, nor how to map each shared-nothing instance to CPU cores.

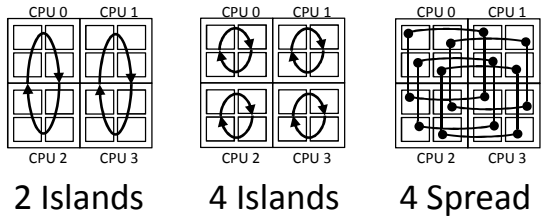

Figure 4: Different shared-nothing configurations on a four-socket four-core machine.

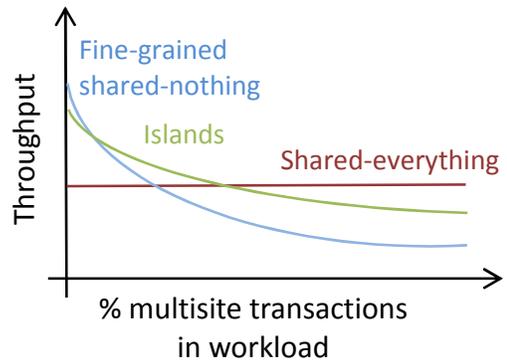

Figure 5: Performance of various deployment configurations as the percentage of multisite transactions increases.

Figure 4 illustrates three different shared-nothing configurations. The two left-most configurations, labeled *"2 Islands"* and *"4 Islands"*, dedicate different number of cores per instance, but, for the given size, minimize the NUMA effects as much as possible. Computation within an instance is done in close cores. The third configuration, *"4 Spread"* has the same size per instance as *"4 Islands"*; however, it does not minimize the NUMA effects, as it forces communication across sockets when it is strictly not needed. The first two configurations are islands in our terminology, where an island is a shared-nothing configuration where each shared-nothing instance is placed in a topology-aware manner. The third configuration is simply a shared-nothing configuration. As hardware becomes more parallel and more heterogeneous the design space over the possible shared-nothing configurations increases, and it is harder to determine the optimal deployment.

On top of the hardware complexity, we have to consider that the cost of a transaction in a shared-nothing environment also depends on whether this transaction is *local* to a database instance or *distributed*. A transaction is local when all the required data for the transaction is stored in a single database instance. A transaction is distributed when multiple database instances need to be contacted and slow distributed consensus protocols (such as two-phase commit) need to be employed. Thus, the throughput also heavily depends on the workload, adding another dimension to the design space and making the optimal deployment decision nearly "black magic." [6]

An oversimplified estimation of the throughput of a shared-nothing deployment as a function of the number of distributed transactions is given by the following. If $T_{local}$ is the performance of the shared-nothing system when each instance executes only local transactions, and $T_{distr}$ is the performance of a shared-nothing deployment when every transaction requires data from more than one database instances, then the total throughput $T$ is:

$$T = (1 - p) * T_{local} + p * T_{distr}$$

where $p$ is the fraction of distributed transactions executed.

In a shared-everything configuration all the transactions are local ($p_{SE} = 0$). On the other hand, the percentage of distributed transactions on a shared-nothing system depends on the partitioning algorithm and the system configuration. Typically, shared-nothing configurations of larger size execute fewer distributed transactions, as each database instance contains more data. That is, a given workload has a set of transactions that access data in a single logical site, and transactions that access data in multiple logical sites, which we call *multisite transactions*. A single database instance may hold data for multiple logical sites. In that case, multisite transactions can actually be physically local transactions, since all the required data reside physically in the same database instance. Distributed transactions are only required for multisite transactions whose data reside across different physical database instances. Assuming the same partitioning algorithm is used (e.g. [12, 27]), then the more data a database contains the more likely for a transaction to be local.

Given the previous reasoning one could argue that an optimal shared-nothing configuration consists of a few coarse-grained database instances. This would be a naive assumption as it ignores the effects of hardware parallelism and variable communication costs. For example, if we consider the contention, then the cost of a (local) transaction of a coarse-grained shared-nothing configuration $C_{coarse}$ is higher than the cost of a (local) transaction of a very fine-grained configuration $C_{fine}$, because the number of concurrent contenting threads is larger. That is, $T_{coarse} < T_{fine}$. If we consider communication latency, then the cost of a topology-aware islands configuration $C_{islands}$ of a certain size is lower than the cost of a topology-unaware shared-nothing configuration $C_{naive}$. That is, $T_{islands} < T_{naive}$.

As a result, this paper makes the case for *OLTP Islands*, which are hardware topology- and workload-aware shared-nothing deployments. Figure 5 illustrates the expected behavior of Islands, shared-everything, and finer-grained shared-nothing configurations as the percentage of multisite transactions in the workload increases. Islands exploit the properties of modern hardware by exploring the sets of cores that communicate faster with each other. Islands are shared-nothing designs, but partially combine the advantages of both shared-everything and shared-nothing deployments. Similarly to a shared-everything system, Islands provide robust performance even when transactions in the workload vary slightly. At the same time, performance on well-partitioned workloads should be high, due to less contention and avoidance of higher-latency communication links. Their performance, however, is not as high as a fine-grained shared-nothing system, since each node has more worker threads operating on the same data. At the other side of the spectrum, the performance of Islands will not deteriorate as sharply as a fine-grained shared-nothing under the presence of e.g. skew.

## 5. EXPERIMENTAL SETUP

In the following sections we perform a thorough evaluation of the benefits of various deployment strategies under a variety of workloads on two modern multisocket multicore machines, one with four sockets of 6-core CPUs and one with eight sockets of 10-core CPUs [7].

---

[6] Explaining, among other reasons, the high compensation for skilled database administrators.

Table 2: Description of the machines used.

| Machine | Description |
|---|---|
| Quad-socket | 4 x Intel Xeon E7530 @ 1.86 GHz |
| | 6 cores per CPU |
| | Fully-connected with QPI |
| | 64 GB RAM |
| | 64 KB L1 and 256 KB L2 cache per core |
| | 12 MB L3 shared CPU cache |
| Octo-socket | 8 x Intel Xeon E7-L8867 @ 2.13GHz |
| | 10 cores per CPU |
| | Connected using 3 QPI links per CPU |
| | 192 GB RAM |
| | 64 KB L1 and 256 KB L2 cache per core |
| | 30 MB L3 shared CPU cache |

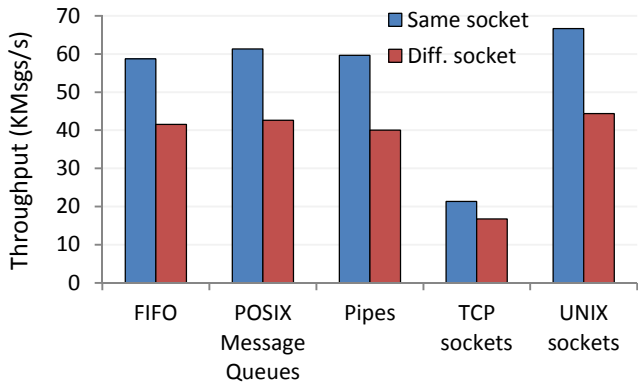

Figure 6: Throughput of message exchanging (in thousands of messages exchanged per second) for a set of inter-process communication mechanisms. Unix domain sockets are the highest performing.

**Hardware and tools.** Table 2 describes in detail the hardware used in the experiments. We disable HyperThreading to reduce variability in the measurements. The operating system is Red Hat Enterprise Linux 6.2 (kernel 2.6.32). In the experiment of Section 7.4, we use two 146 GB 10kRPM SAS 2,5" HDDs in RAID-0.

We use Intel VTune Amplifier XE 2011 to collect basic micro-architectural and time-breakdown profiling results. VTune does hardware counter sampling, which is both accurate and light-weight. Our database system is compiled using GCC 4.4.3 with maximum optimizations. In most experiments, the database size fits in the aggregate buffer pool size. As such, the only I/O is due to the flushing of log entries. However, since the disks are not capable of sustaining the I/O load, we use memory mapped disks for both data and log files. Overall, we exercise all code paths in the system and utilize all available hardware contexts.

**IPC mechanisms.** The performance of any shared-nothing system heavily depends on the efficiency of its communication layer. Figure 6 shows the performance in the quad-socket machine of various inter-process communication (IPC) mechanisms using a simple benchmark that exchanges messages between two processes, which are either located in the same

---

[7]For more details see http://www.supermicro.com/manuals/motherboard/7500/X8OBN-F.pdf

---

CPU socket or in different sockets. Unix domain sockets achieve the highest performance and are used throughout the remaining evaluation.

### 5.1 Prototype System

In order to evaluate the performance of various shared-nothing deployments in multisocket multicore hardware, we implemented a prototype shared-nothing transaction processing system on top of the Shore-MT [18] storage manager. We opted for Shore-MT as the underlying system since it provides near linear scalability on single multicores machines. Shore-MT is the improved version of the SHORE storage manager, originally developed as an object-relational data store [10]. Shore-MT is designed to remove scalability bottlenecks, significantly improving Shore's original single-thread performance. Its performance and scalability are at the highest end of open-source storage managers [18].

Shore-MT is originally a shared-everything system. Therefore, we extended Shore-MT with the ability to run in shared-nothing configurations, by implementing a distributed transaction coordinator using the standard two-phase commit protocol.

Shore-MT includes a number of state-of-the-art optimizations for local transactions, such as speculative lock inheritance [17]. We extended these features for distributed transactions, providing a fair comparison between the execution of local and distributed transactions.

### 5.2 Workload and Experimental Methodology

In our experiments, we vary the number of instances (i.e. partitions) of the database system. Each instance runs as a separate process. In all experiments, the total amount of input data is kept constant and the data is range-partitioned across all instances of the system. For every experiment, with the exception of Section 7.4, we use small dataset with 240,000 rows ($\sim$ 60 MB). We show results using different database configurations, but we always use the same total amount of data, processors, and memory resources. Only the number of instances and the distribution of resources across instances changes.

We ensure that each database instance is optimally deployed. That is, each database process is bound to the cores within a single socket (minimizing NUMA effects) when possible, and its memory is allocated in the nearest memory bank. As noted in Section 3, allowing the operating system to schedule processes arbitrarily leads to suboptimal placement and thread migration, which degrades performance.

The configurations on the graphs are labeled with "NISL" where N represents the number of instances. For instance, 8ISL represents the configuration with 8 database instances, each of which has 1/8th of the total data and uses 3 processor cores (the machine has 24 cores in total). The number of instances is varied from 1 (i.e. a shared-everything system) to 24 (i.e. a fine-grained shared-nothing system). Optimizations are also applied to particular configurations whenever possible: e.g. fine-grained shared-nothing allows certain optimizations to be applied. Optimizations used are noted along the corresponding experimental results.

We run two microbenchmarks. The first consists of read-only transactions that retrieve N rows. The second consists of read-write transactions updating N rows. For each microbenchmark, we run two types of transactions:

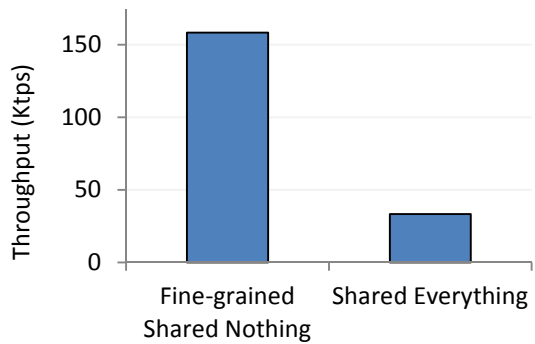

**Figure 7: Running the `TPC-C` benchmark with only local transactions. Fine-grained shared-nothing is 4.5x faster than shared everything.**

- **Local transactions**, which perform its action (read or update) on the $N$ rows located in the local partition;

- **Multisite transactions**, which perform its action (read or update) on one row located in the local partition while remaining $N-1$ rows are chosen uniformly from the whole data range. Transactions are distributed if some of the input rows happen to be located in remote partitions.

## 6. CASES FAVORING FINE-GRAINED PARTITIONING

This section presents two cases where fine-grained shared-nothing configurations outperform coarser-grained shared-nothing configurations as well as shared-everything.

### 6.1 Perfectly Partitionable Workloads

If the workload is perfectly partitionable then fine-grained shared-nothing provides better performance. An example is shown on Figure 7, obtained using the quad-socket machine, which compares the performance of the shared-everything version of Shore-MT with the fine-grained shared-nothing version of Shore-MT with 24ISL. Both systems run a modified version of the `TPC-C` benchmark [34] Payment transaction, where all the requests are local and, hence, the workload is perfectly partitionable on `Warehouse`s. The fine-grained shared-nothing configuration outperforms shared-everything by 4.5x, due in large part to contention on the Warehouse table in the shared-everything case. Experiments with short-running microbenchmarks in later sections, however, do not show such a large difference between shared-everything and shared-nothing. This is because of the larger number of rows in the table being accessed, which implies lower contention on a particular row.

### 6.2 Read-only Workloads

Fine-grained shared-nothing configurations are also appropriate for read-only workloads. In the following experiment we run microbenchmark with local transactions that retrieve 10 rows each. We test multiple configurations ranging from 24ISL to 1ISL in the quad-socket machine. The configuration 24ISL is run without locking or latching.

Figure 8 (left) shows that the fine-grained shared-nothing configurations, whose instances have fewer threads, make better utilization of the CPU. Single-threaded instances, apart from not communicating with other instances, have simpler execution leading to shorter code paths, which decreases the number of instruction misses. On the other hand, instances that span across sockets have a much higher percentage of stalled cycles (shown in the middle of Figure 8), in part due to the significant percentage of—expensive—last-level cache (LLC) misses. Within the same socket, smaller instances have higher ratio of instructions per cycle due to less sharing between cores running threads from the same instance, as shown on the Figure 8 (right).

## 7. CHALLENGES FOR FINE-GRAINED PARTITIONING

A significant number of real life workloads cannot be partitioned in a way that transactions access a single partition. Moreover, many workloads contain data and access skews, which may also change dynamically. Such workloads are more challenging for systems that use fine-grained partitioning and coarser-grained shared-nothing configurations provide a robust alternative.

### 7.1 Cost of Distributed Transactions

Distributed transactions are known to incur a significant cost, and this problem has been the subject of previous research, with e.g. proposals to reduce the overhead of the distributed transaction coordination [20] or to determine an initial optimal partitioning strategy [12, 27]. Our experiment, shown in Figure 9, corroborates these results. We run two microbenchmarks whose transactions read and update 10 rows respectively on the quad-socket machine. As expected, the configuration 1ISL (i.e. shared-everything) is not affected by varying the percentage of multisite transactions. However, there is a drop in performance of the remaining configurations, which is more significant in the case of the fine-grained one. The following experiments further analyze this behavior.

#### 7.1.1 Read-only Case: Overhead Proportional to the Number of Participating Instances

Figure 10 (upper left) represents the costs of a local read-only transaction in various database configurations and as the number of rows retrieved per transaction increases. The results are obtained on the quad-socket machine. The 24ISL configuration runs with a single worker thread per instance, so locking and latching are disabled, which leads to roughly 40% lower costs than the next best configuration, corroborating previous results [15].

The costs of multisite read-only transactions (Figure 10, upper right) show the opposite trend from the local read-only transactions. In the local case, the costs of a single transaction rise as the size of an instance grows. In the multisite case, however, the costs decrease with the size of an instance. This is due to a decrease in the number of instances participating in the execution of any single transaction. The exception is the shared-everything configuration, which has higher costs due to inter-socket communication, as discussed in Section 6.

#### 7.1.2 Update Case: Additional Logging Overhead Is Significant

The lower left plot of Figure 10 describes the costs of the update microbenchmark with local transactions only, on the quad-socket machine. The cost of a transaction increases with the number of threads in the system, due to contention on shared data structures. As in the read-only case, the 24ISL configuration runs without locks or latches and hence, has lower costs.

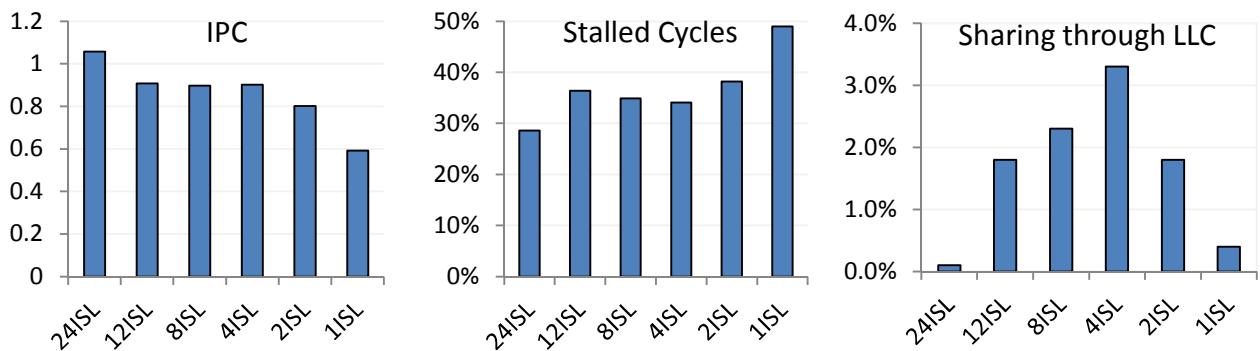

Figure 8: Microarchitectural data for different deployments: instructions per cycle (left), percentage of stalled cycles (middle) and percentage of cycles when data is shared between cores on the same socket (right). IPC is much higher for smaller instances.

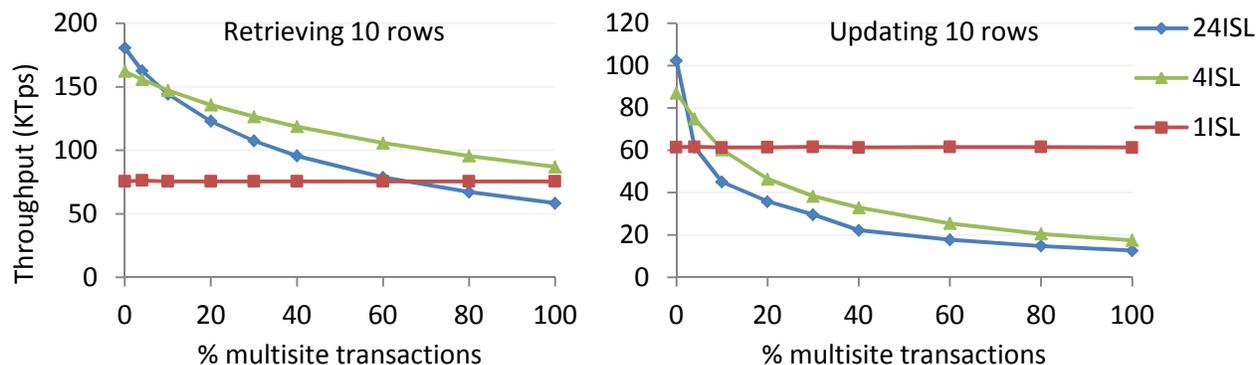

Figure 9: Performance as the number of distributed transactions increases. While shared-everything remains stable, performance of share-nothing configurations decreases.

Multisite shared-nothing transactions (Figure 10, lower right) are significantly more expensive than their local counterparts. This is due to the overhead associated with distributed transactions and to the (mandatory) use of locking. Any configuration that requires distributed transactions is more expensive than the shared-everything configuration.

### 7.1.3 Profiling

To characterize the overhead of inter-process communication costs in relation to the remaining costs of a distributed transaction, we profile the execution of a set of read-only and update transactions on the quad-socket machine, using the 4ISL configuration. Figure 11 plots time breakdown for the lightweight transaction which reads or updates 4 rows. The messaging overhead is high in the read-only case, although it has a constant cost per transaction. The relative cost of communication can be seen by comparing the 0% multisite (i.e. local transactions only) and the 100% multisite bars. Although distributed transactions require exchange of twice as many messages in the update case, this overhead is comparatively smaller because of additional logging, as well as increased contention which further increase the cost of a transaction.

## 7.2 Increasing Hardware Parallelism

Hardware parallelism as well as communication variability will likely continue to increase in future processors. Therefore, it is important to study the behavior of alternative database configurations as hardware parallelism and communication variability grows. In Figure 12, we run the microbenchmark which reads (left) or updates (right) 10 rows with fixed percentage of multisite transactions to 20%, while the number of cores active in the machine is increased gradually. Results are shown for both the quad-socket and the (more parallel and variable) octo-socket machine.

The shared-nothing configurations scale linearly, with $CG$ (coarse-grained shared-nothing) configuration being competitive with the best case across different machines and across different levels of hardware parallelism. The configuration labeled $SE$ (shared-everything) does not scale linearly, particularly on the machine with 8 sockets. In the $SE$ configuration, there is no locality when accessing the buffer pool, locks, or latches. To verify the poor locality of $SE$, we measured the QPI/IMC ratio, i.e. the ratio of the inter-socket traffic over memory controller traffic. A higher QPI/IMC ratio means the system does more inter-socket traffic while reading (i.e. processing) less data overall: it is less NUMA-friendly. The QPI/IMC ratio for the experiment with read-only workload on octo-socket server using all 80 cores is 1.73 for $SE$, 1.54 for $CG$, and 1.52 for $FG$. The $FG$ and $CG$ configurations still have a relatively high ratio due to multisite transactions but, unlike $SE$, these consist of useful work. When restricting all configurations to local transactions only, we observe a steady data traffic of 100 Mb/s on the inter-socket links for $FG$ and $CG$ (similar to the values observed when the system is idle), while $SE$ exceeds 2000 Mb/s. Clearly, to scale the $SE$ configuration to a larger number of cores, data locality has

1454

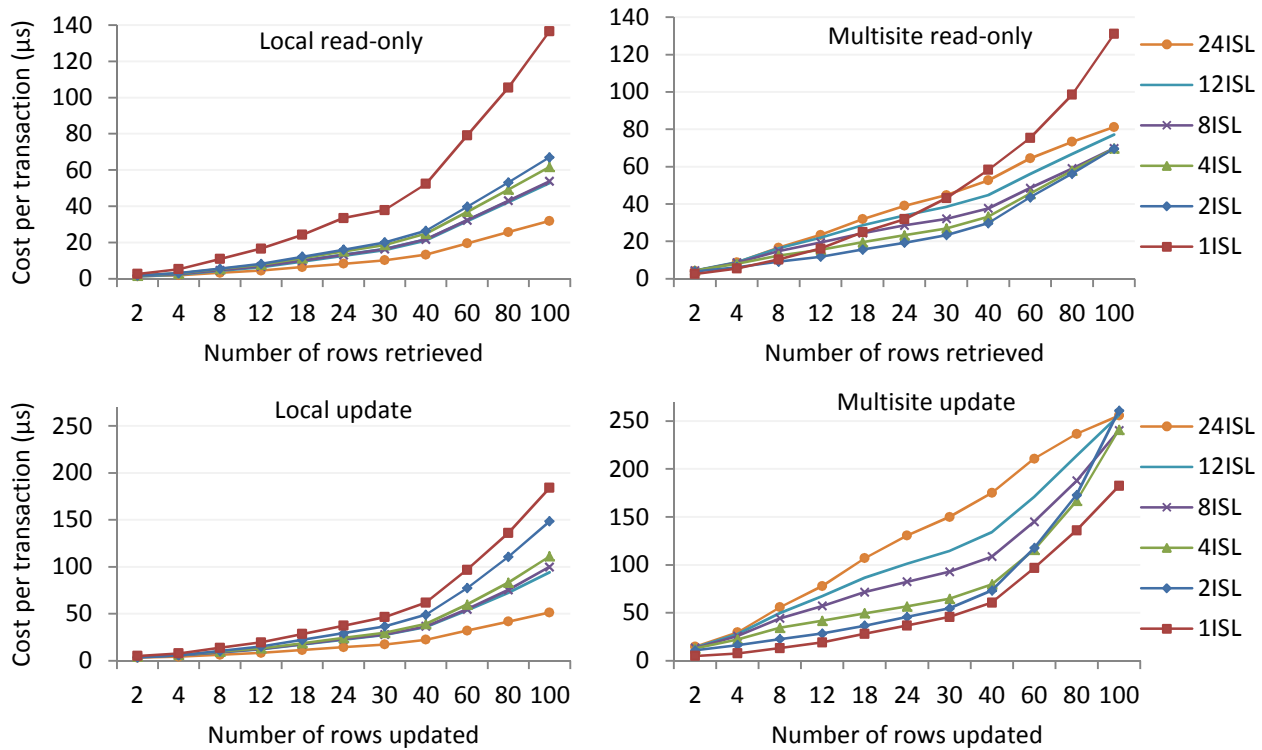

Figure 10: Cost of local and multisite transactions in read-only and update microbenchmarks. Coarse-grained shared-nothing has more robust performance compared to fine-grained shared-nothing and shared-everything.

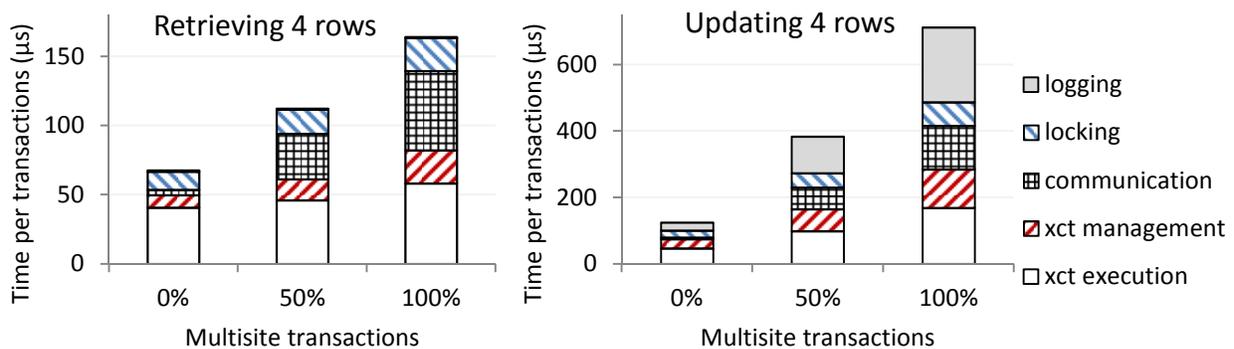

Figure 11: Time breakdown for a transaction that retrieves (left) or updates (right) 4 rows. The cost of communication dominates in the cost of distributed transaction in the read-only case, while in the update case overheads are divided between communication and additional logging.

to be increased. Additionally, one of the main reasons for poor performance of *SE* configuration is high contention on locks and latches. Using partitioned shared-everything designs with data-oriented execution can significantly improve locality of accesses and remove or minimize the overheads coming from lock and latch managers [25, 26].

## 7.3 Tolerance to Skew

In many real workloads, skews on data and requests, as well as dynamic changes are the norm rather than the exception. For example, many workloads seem to follow the popular 80-20 distribution rule, where the 80% of requests accesses only the 20% of the data. This subsection describes experiments with workloads that exhibit skew.

The following microbenchmark reads or updates two rows chosen with skew over the whole data range. We use Zipfian distribution, with different skew factors $s$, shown on the x-axis of Figure 13. The figures show the throughput for varying percentages of multisite transactions. We employ similar optimizations as described in 7.1.1 and 7.1.2.

Skew has a dramatic effect on the performance of the different configurations. For shared-everything, heavily skewed workloads result in a significant performance drop due to increased contention. This effect is apparent particularly in the update case. When requests are not strongly skewed, shared-everything achieves fairly high performance in the update microbenchmark, mainly due to optimized logging, which significantly improves the performance of short read-write transactions [19]. In coarser-grained islands, the increased load due to skewed accesses is naturally distributed among all worker threads in the affected instance. With fine-grained instances, which have a single worker thread, the additional load cannot be divided and the most loaded instance becomes a bottleneck. Furthermore, as the skew increases to the point where all remote requests go to a single instance, the throughput of other instances also drops as they cannot complete transactions involving the overloaded instance.

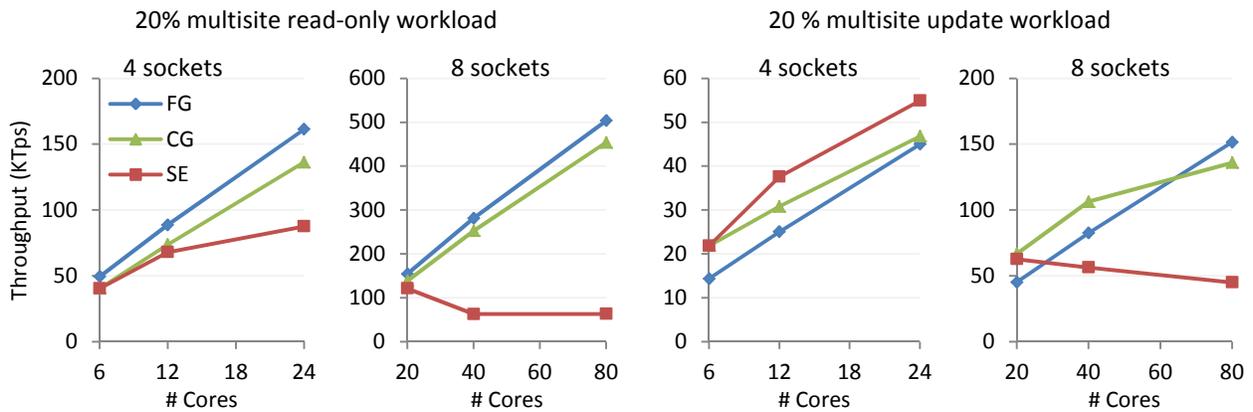

Figure 12: Performance of alternative configurations as the hardware parallelism increases. Coarser-grained shared-nothing provides an adequate compromise between performance and predictability.

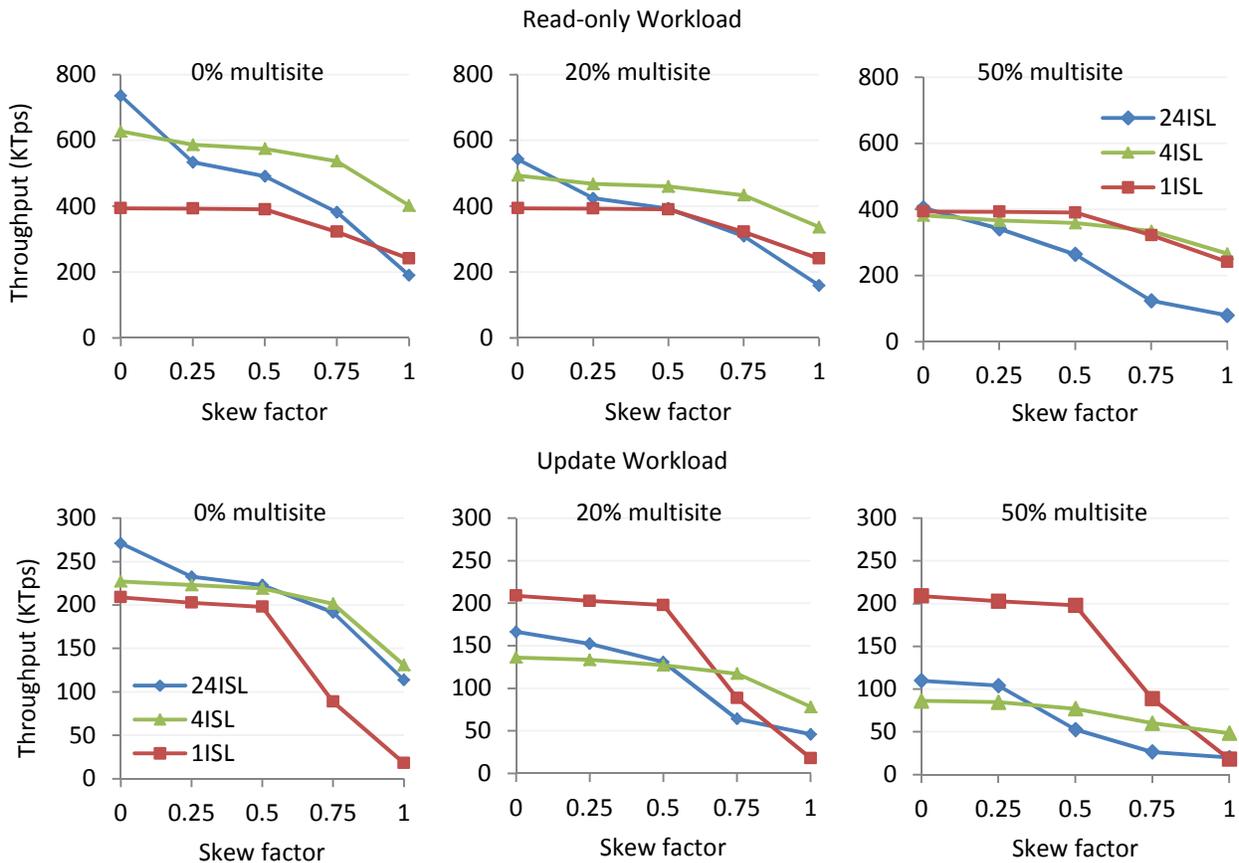

Figure 13: Performance of read-only (top) and update (bottom) workloads with skewed accesses. As skew increases, shared-everything suffers from increased contention, while fine-grained shared-nothing suffers from a highly-loaded instance that slows others. Coarse-grained shared-nothing configuration cope better with a highly loaded instances, due to multiple internal threads.

Overall, coarse-grained shared-nothing configurations exhibit good performance in the presence of skewed requests, as they suffer less from increased contention and are more resistant to load imbalances.

### 7.4 Increasing Database Size

Although main memory sizes in modern servers continue to grow, there are many workloads that are not main memory resident and rely on disk-resident data. To evaluate various database configurations on growing dataset sizes, we gradually increase the number of rows in the dataset from 240,000 to 120,000,000 (i.e. from 60 MB to 33 GB). Contrary to

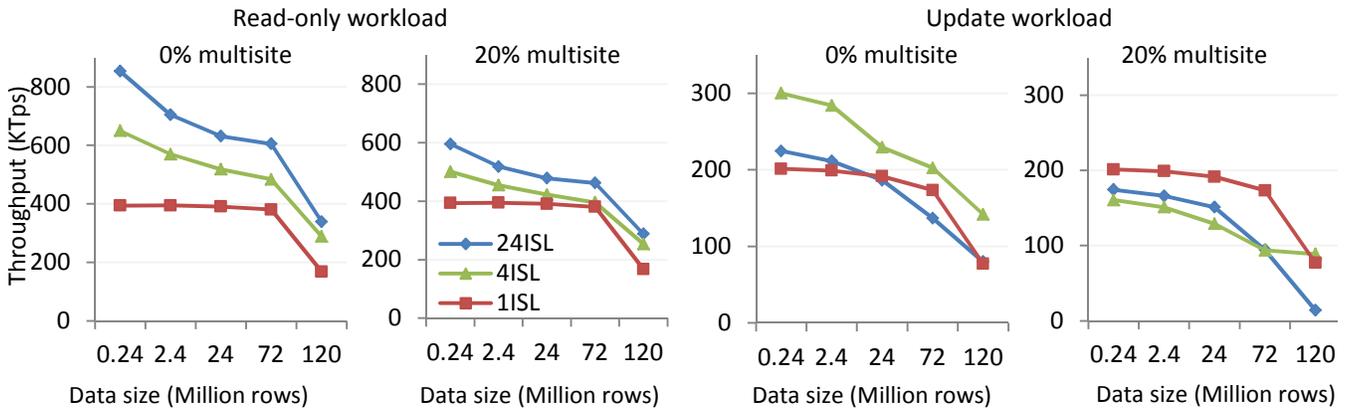
Figure 14: Performance of the various configurations on workloads, as we gradually increase the database size from almost cache-resident to I/O-resident.

previous experiments, we placed the database on two hard disks configured as a RAID stripe. We use a 12 GB buffer pool, so that smaller datasets completely fit in the buffer pool. In the shared-nothing configurations, the buffer pool is proportionally partitioned among instances, e.g. in the 4ISL case each instance has 3 GB buffer pool. We run read and update microbenchmarks with two rows accessed and 0% and 20% multisite transactions.

In Figure 14, we plot the performance of the read-only microbenchmark on the left-hand side and the update microbenchmark on the right-hand side as the number of rows in the database grows. For the smaller dataset, shared-nothing configurations exhibit very good performance as a significant part of the dataset fits in processor last-level caches. Since the instances do not span multiple sockets, there is no inter-socket traffic for cache coherence. As data sizes increase, the performance of shared-nothing configurations decrease steadily, since smaller portions of the data fit in the caches. Finally, when the dataset becomes larger than the buffer pool, the performance drops sharply due to disk I/O. These effect are less pronounced when the percentage of multisite transaction is higher, since the longer latency data accesses are overlapped with the communication.

## 8. CONCLUSIONS AND FUTURE WORK

Modern multisocket multicore servers are characterized by abundant hardware parallelism and variable communication latencies. This non-uniformity has an important impact on OLTP databases and neither traditional shared-everything configurations, nor newer shared-nothing designs, are an optimal choice for every class of OLTP workloads on modern hardware. In fact, our experiments show that no single optimal configuration exists: the ideal configuration is dependent on the hardware topology and workload, but the performance and variability between alternative configurations can be very significant, encouraging a careful choice. There is, however, a common observation across all experiments: **the topology of modern servers favors a configuration we call Islands**, which groups together cores that communicate quicker, minimizing access latencies and variability.

We show that **OLTP Islands provide robust performance under a variety of scenarios.** Islands, being topology and workload-aware, provide some of the performance gains of shared-nothing databases while being more robust to changes in the workload than shared-nothing. Their performance under heavy skews and multisite transactions also suffers, but overall, Islands are robust under the presence of moderate skews and multisite transactions.

The challenge, then, is to determine the ideal island, out of the many possible islands. A straightforward choice is to use **CPU sockets as a rule-of-thumb island.** Islands group hardware cores that communicate faster, which is easily achieved by dimensioning and placing islands to match hardware sockets.

As for previous approaches, our experiments corroborate previous results in that **shared-everything OLTP provides stable but non-optimal performance.** Shared-everything databases are robust to skew and/or updates in their workloads. However, their performance is not optimal and in many cases, significantly worse than the ideal configuration. In addition, **shared-everything OLTP is likely to suffer more on future hardware.** As the hardware parallelism continues to increase, it becomes increasingly important to make shared-everything databases NUMA-aware. Also, **extreme shared-nothing OLTP is fast but sensitive to the workload.** Extreme shared-nothing databases, as advocated by systems such as H-Store, provide nearly optimal performance if the workload is perfectly partitionable. Shared-nothing databases, however, are sensitive to skew and multisite transactions, particularly in the presence of updates.

Future work will focus on determining the ideal size of each island automatically for the given hardware and workload. Moreover, in clustered databases, shared-cache shared-disk designs [23] allow database instances to share buffer pools, avoiding accesses to the shared-disk. Studying the performance of shared-disk deployments within a single multisocket multicore node is also part of our future plans. Scaling-out OLTP across multiple machines is an orthogonal problem, but the islands concept is likely to be applicable.

## 9. ACKNOWLEDGMENTS

We would like to thank Eric Sedlar and Brian Gold for many insightful discussions and the members of the DIAS laboratory for their support throughout this work. This work was partially supported by a grant from Oracle Labs, an ESF EurYI award, and Swiss National Foundation funds.